# THE DYNAMIC EVOLUTION OF THE POWER EXPONENT IN A UNIVERSAL GROWTH MODEL OF TUMORS

Caterina Guiot[†,1,2], Pier Paolo Delsanto[2,3], Alberto Carpinteri[4], Nicola Pugno[4]

Yuri Mansury[5,*] and Thomas S. Deisboeck[5]

**Affiliations:** [1]Dip. Neuroscience, Università di Torino, Italy; [2]INFM, sezioni di Torino Università e Politecnico, Italy; [3]Dip. Fisica, Politecnico di Torino, Italy; ; [4]Dip. Ing. Strutturale e Geotecnica, Politecnico di Torino, Italy ; [5]Complex Biosystems Modeling Laboratory, Harvard-MIT (HST) Athinoula A. Martinos Center for Biomedical Imaging, Charlestown, MA 02129, USA.

[†]**Corresponding Author:**

Caterina Guiot, Ph.D., Dip. Neuroscienze, 30, C. Raffaello, 10125 Torino

tel: +39.11.670.7710, fax: +39.11.670.7708, e-mail: caterina.guiot@unito.it

*Current Address: Yuri Mansury, Ph.D., Dept. of City and Regional Planning, Cornell University, 213 W. Sibley Hall, Ithaca, NY 14853, USA, e-mail : um10@cornell.edu

**Abstract**

We have previously reported that a universal growth law, as proposed by West and collaborators for all living organisms, appears to be able to describe also the growth of tumors *in vivo*. In contrast to the assumption of a fixed power exponent $p$ (assumed by West et al. to be equal to 3/4), we show in this paper the dynamic evolution of $p$ from 2/3 to 1, using experimental data from the cancer literature and in analogy with results obtained by applying scaling laws to the study of fragmentation of solids. The dynamic behaviour of $p$ is related to the evolution of the fractal topology of neoplastic vascular systems and might be applied for diagnostic purposes to mark the emergence of a functionally *sufficient* (or effective) neo-angiogenetic structure.

## 1. Introduction

In a previous paper we have reported that West et al.'s model (2001) of "universal growth" may also be extended to describe the growth dynamics of in vitro and in vivo (clinical) tumors (Guiot et al., 2003). The model assumes that the incoming rate of energy flow $B$ (average resting metabolic energy) for a living organism is related to its mass $m$ by a power law of the type $B \propto m^p$, with $p = 3/4$. Such a value of the power exponent was firstly proposed by Kleiber (1932) and recently justified by West et al. (1997), who argued that the distribution network (i) branches to reach everywhere in any three-dimensional organism (according to a fractal distribution), (ii) has terminal units (e.g., capillaries or terminal xylem) independent of the body size, and (iii) minimizes the total resistance and hence the energy required to distribute nutrients. Also Banavar et al. (1999) approached the problem of defining the exponent for a general distributive system. They showed that $B$ is expected to scale as $M^{D/(1+D)}$ if the efficiency of the vascular network is maximized, D being the dimensionality of the embedding space. Thus, in a three-dimensional space $p=3/4$ follows from the condition of most efficient space-filling, without having to recur to fractals.

However, the "correct" value of the p exponent remains a controversial issue since other values have also been proposed in the literature. For instance, a recent paper (Dodds et al., 2001) shows that $p=3/4$, which was the classical value considered in the pioneer paper by Kleiber (1932), does not always yield a significantly better fit for all available data than $p=2/3$, which is based on a simple geometrical scaling of the body surface area available for heat dissipation. Even the trivial assumption *p*=1 has been justified, as the one leading to the simplest hypothesis of direct proportionality between incoming energy flow and mass of the organism.

In our opinion the controversy reflects the rather ambigous formulation of the question. Nature is of course more complex than even the most sophisticated mathematical model. Changing the ingredients of the model affects the prediction of the *p* value and, in turn, may correspond to different phases in the growth of the organism (or tumor). Thus the universality, as defined e.g. in the framework of a recently proposed 'black box' formalism by Hirsekorn and Delsanto (2004), would refer not to a single value of p, but to a suitable range of values. Here in this present contribution we suggest that *p* changes dynamically in the range 2/3 to 1, reflecting the different developmental stages of the vascular network and, more specifically, the development of angiogenesis. Our conjecture is supported by the prediction of similar results by Carpinteri and Pugno (2002a) in a completely different context (see next Section) and by several instances of observation of evolution of the fractal cancer topology (Section 3). In Section 4 a mathematical model of growth dynamics with variable p is presented and some results are drawn and discussed in the Conclusions.

**2. Correlation between scaling laws and fractal cancer topology**

In a completely different context Carpinteri and Pugno (2002a) have developed universal scaling laws for energy dissipation during the fragmentation of solids, by assuming a self-similar (i.e., fractal) size distribution of fragments. Their assumption implies a power law such as $N \propto r^{-\overline{D}}$, where N is the number of fragments with size larger than *r*, and $\overline{D}$ is the so-called fractal exponent (a real positive number) of the fragment size distribution. Accordingly they obtain by integration the total surface *S* of fragments, as a function of their total volume *V*, as $S \propto V^{\overline{D}/3}$, with $2 < \overline{D} < 3$.

Likewise, if the biological clusters (fragments) distribution is fractal in nature, and the energy transportation (dissipation) is proportional to their surface, neglecting the variation of density during growth ($V \propto m$) yields the scaling law $B \propto m^p$, with $p = \overline{D}/3$, with $2/3 < p < 1$, as conjectured in the Introduction.

It is interesting to note that, according to the interpretation based on the analysis by Carpinteri and Pugno (2002a), the exponent $p$ should be strongly related to the fractal nature of cancer topology and thus susceptible of independent measurements. The idea of a fractal topology has been proposed in the past by many researchers (see e.g. Baish and Jain, 2000), in particular Baish et al. (1996) have shown that in-vivo estimations of the fractal dimension of planar vascular networks based on the box-counting method, (see Bunde and Havlin, 1994) range between 1 and 2. Starting from the 2D observed exponent stereological methods give an estimate of the corresponding 3D value. From a practical point of view the result is close to the 2D fractal exponent plus one, i.e. $\overline{D}$ would range between 2 and 3. Correspondingly, the value of $p = \overline{D}/3$ is comprised between 2/3 and 1. In particular, in normal tissues and in four different tumor lines, implanted in the dorsal skinfold chamber in immunodeficient mice, Baish et al. (1996) observed a value of 2 (corresponding to $\overline{D} = 2+1=3$ in three-dimension and $p =1$) for normal capillaries, 1.7 ($\overline{D} =2.7$, $p =0.9$) for arteries and vein and 1.88 ($\overline{D} =2.88$, $p =0.96$) for tumor vessels, showing that tumor vasculatures are more chaotic and inefficient than normal capillaries, which are almost uniformly distributed. These observations are in agreement with our conjecture and, more specifically, with the prediction of a scaling exponent of 2 for the 'space-filling' growth model, 1.71 for the 'diffusion limited aggregation' model and 1.90 for the 'invasion percolation' model (Baish et al., 1996). Thus tumor vascularization does not fully satisfy the condition of a 'space-filling network', assumed by West et al. (1997), but could perhaps be better described by an 'invasion percolation' model. Accordingly, normal tissues and tumors differ deeply in their vascular structure and metabolism.

## 3. Evolution of the fractal cancer topology

We report here some instances of observed evolution of the fractal cancer topology. The first work, by Gazit et al. (1997) concerns the vascular system changes during growth of tumors implanted in mice. Both the fractal dimension $\overline{D}$ and the vessel density were monitored on two-dimensional images during normal development from the 6-th to the 12-th day, showing an increase from around 1.6 ($\overline{D}$=2.6 in three-dimensions) to a maximum value of 1.73 ($\overline{D}$=2.73) on the 10-th day, followed by a decrease. Likewise also the vessel density shows a large increase, reaching its maximum on the 11-th day, before it decreases. The authors were able to estimate a nearly linear increase in $\overline{D}$ of 0.06 per day correlated to a nearly linear increase in vessel density of 138/cm$^2$ per day.

In another relevant paper the extraembryonic vascular network of the chick embryo was investigated by Vico et al (1998) with similar methods. They found that the vascular fractal dimension increases continuously from about 1.3 ($\overline{D}$=2.3) by the 60-th hour until about 1.68 ($\overline{D}$=2.68) by the 112-th hour, when a plateau is reached and $\overline{D}$ remains stable at the value of approximately 1.7 (D=2.7). Provided that the angiogenetic process is antagonized with angiostatic factors, the fractal dimensionality of the vascular network has been proven to reflect the observed decrease in branching patterns.

Also Guidolin et al (2003) showed that, after delivering docetaxel to cultured HUVEC cells in Matrigel, the fractal dimension decreases of about 10% from the starting value of 1.20 ($\overline{D}$=2.20). Finally, the already cited paper by Parsons-Wingerter et al. (1998) shows similar effects after angiostatin delivering in the quail chorioallontoic membrane.

## 4. A mathematical model

According to the ontogenetic growth model of West et al. (2001) and its extension to neoplastic growths by Guiot et al (2003), the actual mass $m$ of the tumor and its rate of growth, $dm/dt$, are non-linearly related:

$$\frac{dm}{dt} = a\, m^p \left[1 - \left(\frac{m}{M}\right)^{1-p}\right], \tag{1}$$

where $M$ is the asymptotic value of $m(t)$ and $a$ is a parameter related to the metabolic rate of the particular tumor cell line considered. We have removed the assumption of $p=3/4$ replacing it with $p \in (2/3, 1)$, as conjectured in the Introduction. An unspecified value of $p$ has also been assumed in a paper by Delsanto et al (2004), in which the ontogenetic growth model is tested under controlled conditions of malnourishment and applied mechanical stress. From Eq (1) the universal growth law follows:

$$r = 1 - e^{-\tau}, \text{ with } r = (m/M)^{1-p} \text{ and } \tau = (1-p)bt - \ln(1 - r_0), \tag{2}$$

where $m_0 = m(t=0)$, $r_0 = (m_0/M)^{1-p}$ and

$$b = a\, M^{p-1} \tag{3}$$

For $p=3/4$ the growth predicted by Eq. (2) coincides with the West et al.'s model (Guiot et al., 2003).

Using Eq. (1) it can be shown that $m(t)$ exhibits an inflection point at $t=t'$, corresponding to $m=m'$, which depends on the value of $p$. The simplest way to determine $m'$ is to plot the experimental values of *m* and d*m*/d*t* in a cartesian plane with *m* as abscissa and d*m*/d*t* as ordinate. The curve d*m*/d*t* vs. *m* reaches a maximum at *m=m'*, which can be accurately identified. By assuming a dynamic evolution of the fractal exponent *p=p(t)*, but with a slow rate of change (i.e. neglecting d*p/*d*t* in the derivatives) it follows that:

$$\frac{d\mu}{dt} = b\, \mu^p \left( \frac{1}{p} - \mu^{1-p} \right), \tag{4}$$

where $\mu = \frac{m}{m'}$ and:

$$m' \cong p^{\frac{1}{1-p}} M. \tag{5}$$

From the plot d$\mu$/d*t* vs. $\mu$ we obtain the best fitting values of *b* and of the exponent *p* corresponding to the mass *m'* (and time *t'*). Then *M* and *a* can be immediately computed from Eqs. (5) and (3), respectively. At this point *p* can be evaluated in its dynamical evolution (i.e. for each value of *m* and *t*) by mean of Eq. (4).

As a first example of application of the procedure, we consider the analysis reported by Torres et al. (1995) in which a model for the investigation of angiogenesis is presented. It consists in the implantation of a Lewis's lung carcinoma multicellular tumor spheroid into the dorsal skinfold chamber of mice. The authors monitored both the morphometric parameters of tumor growth and the development of the vascular networks around the tumor focus, using intravital microscopy. By applying our procedure to their data we obtain the results reported in Figure 1. After an initial decrease, *p* starts to grow up to a value around 3/4 when the angiogenetic process is apparently completed. According to Parsons-Wingerter et al. (1998) and Vico et al. (1998), *p* is expected to grow with the vascular density. However, from our plot it appears that *p* starts growing only when the vascular density has already reached a considerable level. Unfortunately, the latter authors do not consider the evolution from its very beginning, yet they also show some delay in the relationship between fractal dimension and vascular density. We may infer that the initial decrease of $p$ is due to the transition from the original (optimal) network to the subsequent pre-angiogenetic structure. A new vascular network needs to be well established before inducing a growth in $p$.

As a second example (Fig. 2) we investigate three out of five cell lines of tumors growing in mice, as reported by Steel (1977). The data from the other two cell lines cannot be used for this procedure since the short duration of the corresponding experimental series does not allow a proper estimate of the tumor mass at the inflection point. Note that, regardless of the cancer type, the power exponent $p$ is observed to change dynamically, i.e., after an initial decrease (related to the implantation process as previously discussed), it eventually rises up to saturation.

## 5. Discussion and concluding remarks

In this paper we have studied the correlation between tumor topology and the scaling exponent $p$, which we have conjectured to vary dynamically in the range (2/3,1). Consequently, we have modified the ontogenetic growth model of West et al. (2001) and its extension to neoplastic growths by Guiot et al (2003), in order to develop a model for the prediction of the dynamic behaviour of $p$ .For the application of the model to the analysis of experimental data (tumor masses $m$) we have considered the plot of $dm/dt$ vs. $m$, which allows in the simplest way to evaluate $p$. We found that, in general, after an initial decrease due to the 'adaptation' of the implanted tumor to the new environment in the avascular phase, $p$ starts increasing. We conjecture that this point marks the switch in the dominant nutrient-replenishment mechanism, from passive diffusion to active perfusion conferred by a then functionally sufficient level of angiogenesis. This transition occurs in the three cell lines investigated by Steel (1997) at an average tumor diameter of 6.6 mm (± 1.6 STD), clearly beyond the threshold of 2-3 mm which, according to Folkman (1971), should prompt the onset of angiogenesis. Values of *p* beyond 0.75 may suggest that active perfusion is complemented by other supply mechanisms, such as passive diffusion, when vascular density approaches its plateau. For the data analyzed here, this dynamic *p*-behaviour appears to be independent of the *in vivo* cancer type. It is also interesting to note that the time at which p starts growing (supposedly following the onset of angiogenesis, i.e., after the neovascular network already improves distribution efficiency) ranges between 5.3 and 14.2 days after implantation of the tumor. After rescaling to the dimensionless time $\tau$ defined in Eq. 2, this temporal interval falls into a much narrower range (from about 0.21 to 0.39). This $\tau$ range might be further reduced with a proper analysis of the implantation mechanism.

Based on the presented results, we argue that the scaling exponent *p* shows distinct dynamic patterns *in vivo* and that a monitoring of *p* may be of interest for diagnostic or therapeutic purposes if the correspondence of the minimum of p with the emergence of a functionally *sufficient* neoangiogenetic structure is confirmed. More specific experiments, including measurements of tumor volume and vascular density and fractality would be desirable, given the important implications of our assumptions for cancer research.

**Acknowledgements**

This work was supported in part by the National Institutes of Health (CA 085139 and CA 113004) and by the Harvard-MIT (HST) Athinoula A. Martinos Center for Biomedical Imaging and the Department of Radiology at Massachusetts General Hospital. We also wish to thank Drs. M. Griffa and P.G. Degiorgis for useful discussions.

**Figure Caption**

Figure 1. Prediction according to our model of the scaling exponent *p* vs. time, based on data from Torres et al. (1995) referring to tumors implanted in CB6 mice. The corresponding values of the vascular density are also reported.

Figure 2. Prediction according to our model of the scaling exponent *p* vs. time, based on data from Steel (1977) referring to three different tumor cell lines implanted in mice.

Figure 1

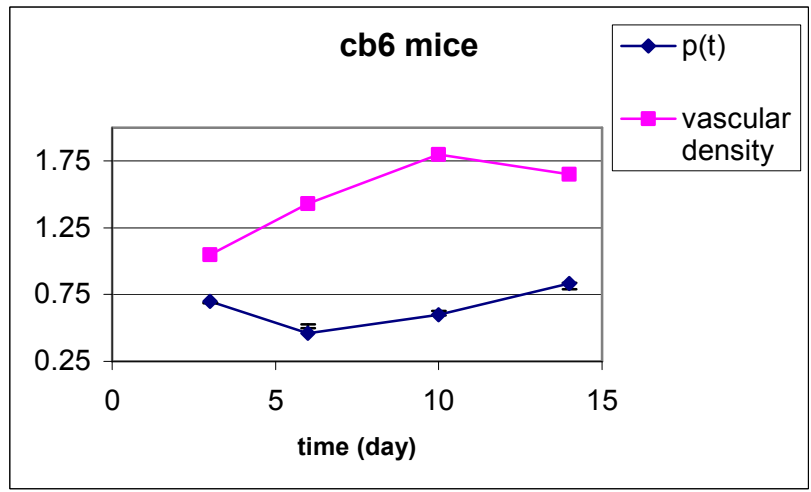

Figure 2

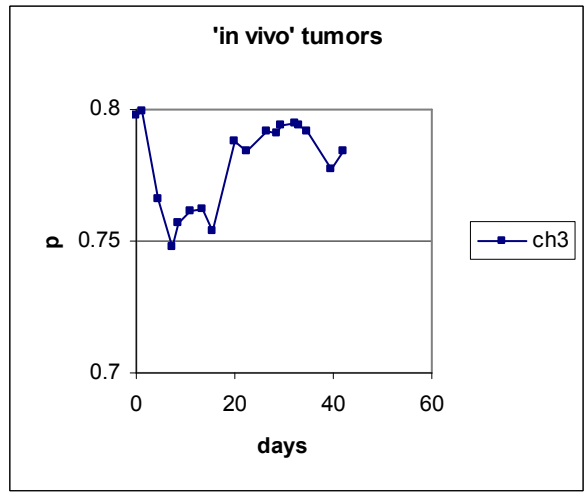

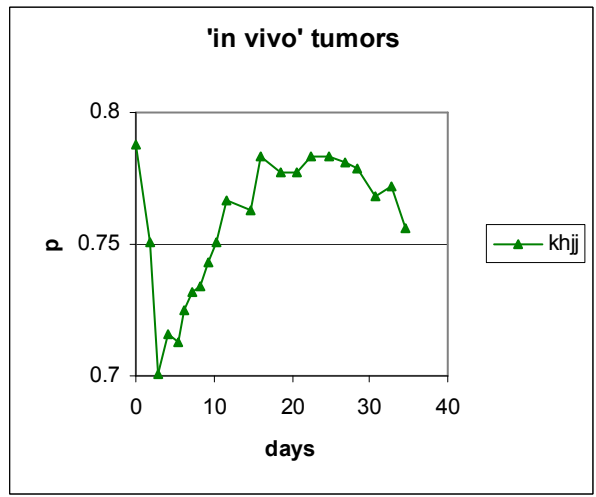

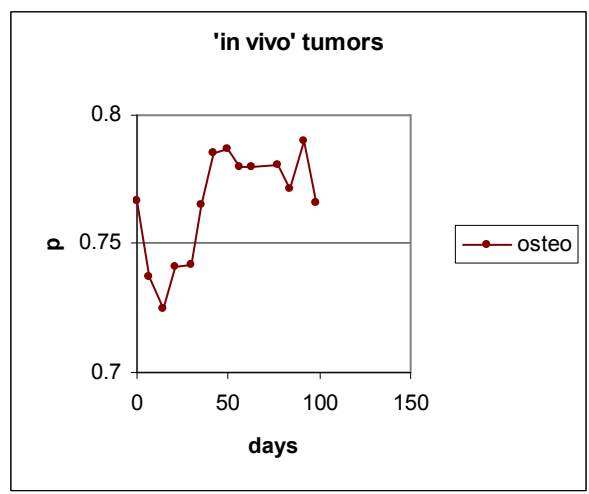